\documentclass[12pt]{article}

\usepackage[margin=1in]{geometry}

\usepackage{newtxtext,newtxmath}

\usepackage{graphicx}
\usepackage{array}
\usepackage{booktabs}
\usepackage{rotating}
\usepackage{float}
\usepackage{xcolor}
\usepackage{amsmath, amsfonts}    
\usepackage{float}
\usepackage{indentfirst}
\usepackage{subcaption}
\usepackage{adjustbox}
\usepackage[super,sort&compress,numbers]{natbib}
\usepackage{url}
\usepackage[colorlinks=true, allcolors=blue]{hyperref}

\begin{document}

\title{Comparing Tobit and Two-Part Hurdle Models for Semi-Continuous Longitudinal Data with an Application to Clonal Hematopoiesis}

\author{%
\parbox{\textwidth}{\centering
\small
Sumaja Bandreddi$^{1,\dagger}$, Pei Zhang$^{1,\dagger,\ddagger}$, Rebecca L. Kelly$^{2}$, Mitchell J. Machiela$^{2}$, Paul S. Albert$^{1,\ast}$\\[3pt]
$^{1}$\textit{Biostatistics Branch, Division of Cancer Epidemiology and Genetics, National Cancer Institute, 9609 Medical Center Drive, Rockville, Maryland 20850, United States}\\[2pt]
$^{2}$\textit{Integrative Tumor Epidemiology Branch, Division of Cancer Epidemiology and Genetics, National Cancer Institute, 9609 Medical Center Drive, Rockville, Maryland 20850, United States}\\[3pt]
$^{\dagger}$These authors contributed equally to this work.\\[2pt]
$^{\ddagger,\ast}$Correspondence: Dr. Paul S. Albert (\texttt{albertp@mail.nih.gov}) and Dr. Pei Zhang (\texttt{pei.zhang@nih.gov})
}}

\date{}

\maketitle

\begin{abstract}
{

Zero-inflated nonnegative continuous (semi-continuous) longitudinal data frequently arise in biomedical studies where outcomes consist of a mixture of excess zeros and positive continuous measurements. Two widely used approaches for analyzing such data are mixed-model versions of  Tobit and the two-part hurdle models. The Tobit assumes a latent regression model that is censored below a specified threshold, while the hurdle separately models the continuous positive outcomes and the binary indicator of being positive. The choice between these models has rarely been systematically discussed, and inappropriate model choice may lead to biased estimation  and misleading scientific interpretations. 

In this paper, we derive rigorous mathematical conditions under which the two models are equivalent and show that the Tobit  can be viewed as a special case of the hurdle model when the link function for the binary process is probit. Based on simulation studies, we found that the hurdle is more flexible and robust than the Tobit model. On the other hand, if the assumptions of the Tobit are met (which  are  difficult to verify in practice), this model is easier to interpret since it does not require distinct inferences on both the continuous and binary processes.  We therefore recommend that the Tobit model only be used  when these assumptions are scientifically plausible and empirically supported; otherwise, the hurdle model is preferable. 

We applied both models to study the dynamics of somatic mosaicism using longitudinal clonal fraction measurements from the Prostate, Lung, Colorectal,and Ovarian (PLCO) study data while accounting for excess zero values. Estimates obtained from the Tobit and hurdle models were broadly consistent with those from a standard linear model that ignored zero inflation. These findings provide additional support for previously reported associations in studies of clonal hematopoiesis across different types of mosaic chromosomal alterations. 
}

\end{abstract}
\vspace{0.4em}

\section{Introduction}
\label{sec1}

Semi-continuous or zero-inflated non-negative data are frequently encountered in biomedical research, where the outcome has a mix of zeros and positive continuous values. Examples include measures of health status \citep{austin2000use}, acupuncture clinical trial \citep{albert2005modelling},  medical cost data \citep{tian2007two},  single cell gene expression \citep{finak2015mast}, microbiome data \citep{chen2016two} and alcohol binge drinking \citep{ren2022semi}. Excluding zero observations or analyzing only positive values may yield biased estimates and potentially misleading scientific conclusions \citep{su2009bias}.

Two classical models are commonly used for such data. The Tobit model, originally proposed by \citet{tobin1958estimation}, treats zeros as censored observations from an underlying latent continuous variable subject to a detection limit, assuming that zeros and positive values arise from the same latent Gaussian process \citep{hughes1999mixed, proust2014joint}. In contrast, the two-part, or hurdle, model introduced by \citet{cragg1971some} separates the occurrence and magnitude processes, with one component modeling whether the outcome is zero or positive and the other modeling the positive outcome conditional on being positive.

We focus on longitudinal zero-inflated non-negative continuous outcomes, hereafter referred to as longitudinal semi-continuous data. Longitudinal Tobit regression has been used for repeated outcomes with floor effects or detection limits by treating zeros as censored realizations from an underlying latent continuous process \citep{twisk2009longitudinal}. By incorporating subject-level random effects, Tobit mixed models account for within-subject dependence. As an alternative, \citet{olsen2001two} introduced a two-part random-effects model that separately models zero occurrence and positive outcome magnitude. Correlated random effects in two-part mixed models allow these two processes to be associated at the subject level, although independence is often assumed for computational convenience. \citet{su2009bias} showed that imposing independence when the random effects are truly correlated can bias regression coefficient estimates. Motivated by these considerations, we study the Tobit random-effects model and the two-part hurdle model with correlated random effects, with particular attention to their formal relationship, assumptions, and practical implications for longitudinal semi-continuous data. 

In this paper, we compare these models in the presence of a detection limit, motivated by prior evidence that the Tobit model can be biased when observed zeros represent true absence rather than censoring \citep{kim2009two,noh2019extended}. By examining their likelihood formulations, we show that the Tobit model can be viewed as a constrained special case of the hurdle model with a probit link. The Tobit model may therefore be overly restrictive, as it attributes zeros to the lower tail of a single latent Gaussian process. In contrast, the hurdle model offers greater flexibility by modeling threshold exceedance separately from the distribution of positive outcomes. This distinction is particularly important for semi-continuous outcomes with a large proportion of zeros, for which the Tobit model may be inappropriate, whereas the hurdle model can continue to perform well \citep{moulton1995mixture}. Our analysis provides a comprehensive comparison of these two approaches and highlights several practical considerations for researchers when selecting and implementing models for semi-continuous data.

The methodological motivation for this work stems from a longitudinal investigation of mosaic chromosomal alterations (mCAs) in the Prostate, Lung, Colorectal, and Ovarian  (PLCO) Cancer Screening Trial reported by \citet{kelly2025}.  
The clonal fraction (CF), which measures the proportion of blood cells carrying a given alteration, is inherently semi-continuous, with many zero observations and continuous positive values. These zeros may reflect either true absence of a clone or values below the assay detection limit. We therefore reanalyze the PLCO Study data using Tobit and hurdle models that explicitly accommodate the semi-continuous longitudinal structure, and compare the resulting covariate effects with those from a positive-value analysis presented by \citet{kelly2025}.

The article is organized as follows. Section~\ref{sec:methods} introduces the Tobit and  hurdle models for semi-continuous longitudinal data, describes their implementation in R, and establishes the theoretical relationship between these modeling frameworks. Section~\ref{sec:application} applies both models to longitudinal clonal hematopoiesis data from the PLCO cohort and compares the resulting estimates across mosaic event types. Section~\ref{sec:simulations} evaluates model performance through simulation studies under correct specification, cross-model fitting, and excess-zero scenarios. Section~\ref{sec:discussion} concludes with methodological findings, practical recommendations, and directions for future research.

\section{Method} \label{sec:methods}

\subsection{Statistical Models}

For subject \(i = 1,\ldots,N\) at visit \(j = 1,\ldots,n_i\), measured at time \(t_{ij}\), let \(y_{ij}\) denote the semi-continuous longitudinal outcome. Such  outcomes are characterized by a mixture of 
a point mass at zero, representing either true zeros or a non-detectable observations, and a continuous distribution of positive values exceeding a detection threshold. For a generic threshold \(L\), define
\[
Z_{ij}(L)=\mathbb{I}(y_{ij}>L),
\]
where \(Z_{ij}(L)=1\) indicates that the observed value exceeds the threshold. For simplicity, we use a common threshold \(L\) for these two models.


\subsubsection{Tobit Model}

The Tobit model was originally proposed to accommodate censored outcome data \citep{tobin1958estimation}. In the longitudinal semi-continuous setting, the model assumes that the observed outcome arises from an underlying latent Gaussian process, with values below a detection threshold recorded as zero or censored at the threshold \citep{amemiya1984tobit,dagne2017joint,liu2019statistical}.

Let \(y_{ij}^{\ast}\) denote the latent outcome for subject \(i\) at visit \(j\). We specify
\begin{equation} \label{eq:tobit_model_latent}
  y_{ij}^*
  =
  \beta_0
  +
  \sum_{k=1}^{K}\beta_k x_{ij}^{(k)}
  +
  b_{i0}
  +
  b_{i1}t_{ij}
  +
  \varepsilon_{ij},
\end{equation}
where \(\beta_0,\ldots,\beta_K\) are fixed-effect coefficients, \(x_{ij}^{(k)}\) denotes the \(k\)th covariate, and \(x_{ij}^{(1)}:=t_{ij}\) may denote time. The subject-specific random effects \(\boldsymbol b_i=(b_{i0},b_{i1})^\top\) are assumed to follow
\[
\boldsymbol b_i \sim N(\boldsymbol 0,\Sigma_b),
\]
with
\[
\Sigma_b =
\begin{pmatrix}
    \sigma^2_{b0} & \rho_b \sigma_{b0}\sigma_{b1} \\
    \rho_b \sigma_{b0}\sigma_{b1} & \sigma^2_{b1}
\end{pmatrix}.
\]
The residual errors satisfy \(\varepsilon_{ij}\sim N(0,\sigma_\varepsilon^2)\) and are assumed independent of the random effects.\\

The observed outcome is generated from the latent outcome as
\begin{equation} \label{eq:tobit_model_observed}
   y_{ij} =
\begin{cases}
0, & \text{if } y_{ij}^* \leq L, \\
y_{ij}^*, & \text{if } y_{ij}^* > L.
\end{cases}
\end{equation}
Thus, zeros are interpreted as left-censored observations arising from the lower tail of a single latent Gaussian process. Under this formulation, the same latent process determines both the probability of exceeding the detection threshold and the magnitude of the observed positive outcome. The fixed-effect coefficients are interpreted on the latent outcome scale.

\subsubsection{Hurdle Model}

In contrast to the Tobit model, the two-part hurdle model separates the process determining whether the outcome exceeds the detection threshold from the process determining the magnitude of positive observations. This formulation is useful when zeros may represent true absence rather than censored values from a latent continuous distribution \citep{mullahy1998much,farewell2017two}. Unlike the Tobit model, the hurdle model makes no assumptions on the outcome distribution below the detection limit (e.g., it does distinguish between true zeros and left censoring).
For longitudinal semi-continuous data, \citet{olsen2001two} extended the two-part model by incorporating subject-specific random effects into both components.

The continuous component models the positive observations conditional on exceeding the threshold:
\begin{equation} \label{eq:hurdle_model_outcome}
 y_{ij}\mid y_{ij}>L
 =
 \gamma_0
 +
 \sum_{s=1}^{S}\gamma_s m_{ij}^{(s)}
 +
 a_{i0}
 +
 a_{i1}t_{ij}
 +
 e_{ij},
\end{equation}
where \(\gamma_0,\ldots,\gamma_S\) are fixed-effect coefficients, \(m_{ij}^{(s)}\) denotes the \(s\)th covariate, and \(m_{ij}^{(1)}:=t_{ij}\) may denote time. The continuous-component random effects \(\boldsymbol a_i=(a_{i0},a_{i1})^\top\) follow
\[
\boldsymbol a_i \sim N(\boldsymbol 0,\Sigma_a),
\]
with
\[
\Sigma_a =
\begin{pmatrix}
    \sigma^2_{a0} & \rho_a \sigma_{a0}\sigma_{a1} \\
    \rho_a \sigma_{a0}\sigma_{a1} & \sigma^2_{a1}
\end{pmatrix}.
\]
The residual errors satisfy \(e_{ij}\sim N(0,\sigma_e^2)\) and are assumed independent of the random effects.

The binary component models the probability of exceeding the detection threshold. Using a probit link, we specify
\begin{align} \label{eq:binary_model}
\begin{split}
    \Pr(y_{ij}>L \mid \boldsymbol c_i)
    & =
    \Pr(Z_{ij}(L)=1\mid \boldsymbol c_i) \\
   &  =
    \Phi\left(
    \alpha_0
    +
    \sum_{s=1}^{S}\alpha_s m_{ij}^{(s)}
    +
    c_{i0}
    +
    c_{i1}t_{ij}
    \right),
    \end{split}
\end{align}
where \(\alpha_0,\ldots,\alpha_S\) are fixed-effect coefficients for the occurrence process, \(\Phi(\cdot)\) is the standard Gaussian cumulative distribution function, and \(\boldsymbol c_i=(c_{i0},c_{i1})^\top\) denotes the binary-component random effects. We assume
\[
\boldsymbol c_i \sim N(\boldsymbol 0,\Sigma_c),
\]
with
\[
\Sigma_c =
\begin{pmatrix}
    \sigma^2_{c0} & \rho_c \sigma_{c0}\sigma_{c1} \\
    \rho_c \sigma_{c0}\sigma_{c1} & \sigma^2_{c1}
\end{pmatrix}.
\]

To allow subject-specific association between the occurrence and magnitude processes, the random effects from the two components may be correlated. Specifically, the joint distribution of \((\boldsymbol a_i^\top,\boldsymbol c_i^\top)^\top\) is assumed Gaussian, with cross-component covariances
\[
\text{cov}(a_{i0},c_{i0})=\rho_0\sigma_{a0}\sigma_{c0}, \qquad
\text{cov}(a_{i1},c_{i1})=\rho_1\sigma_{a1}\sigma_{c1},
\]
\[
\text{cov}(a_{i0},c_{i1})=\rho_2\sigma_{a0}\sigma_{c1}, \qquad
\text{cov}(a_{i1},c_{i0})=\rho_3\sigma_{a1}\sigma_{c0}.
\]
If these cross-component covariances are zero, the occurrence and magnitude processes may still share covariates but have no additional subject-specific association. Specifically, when \(\rho_0 = \rho_1 = \rho_2 = \rho_3 =0\), the hurdle model decomposes into two independent submodels: a linear mixed model for the positive continuous outcomes and a generalized linear mixed model with a probit link for the binary occurrence process. We refer to this specification as the separate model throughout the remainder of the paper.

Although other link functions, such as the logit link, can be used for the binary component \citep{smith2017marginalized}, we focus on the probit link because it makes the relationship between the Tobit and hurdle models more transparent.

\subsubsection{Estimation}

All models were fit within a Bayesian framework using the \texttt{brms} package in R, which interfaces with Stan for posterior computation via Hamiltonian Monte Carlo \citep{burkner2017brms,carpenter2017stan}. This framework allows flexible estimation of longitudinal mixed-effects models with subject-specific random effects and provides posterior inference for all model parameters.

For the Tobit model, left censoring at the detection threshold was implemented using the \texttt{cens()} functionality in \texttt{brms}. Under this formulation, observed zeros were treated as censored realizations from an underlying latent Gaussian process. The model included fixed effects, such as age, smoking status, and their interaction, together with subject-specific random intercepts and random slopes for time.

For the hurdle model, we used a two-part specification. The binary component modeled the probability of exceeding the detection threshold using a Bernoulli distribution with a probit link, whereas the continuous component modeled positive observations using a Gaussian distribution. Both components included the same fixed-effect structure and subject-specific random intercepts and slopes to account for within-subject dependence over time.



The likelihood formulation for the hurdle model assumes that conditional on being above $L$, the continuous outcome is Gaussian. In practical applications, we may transform the continuous variable to satisfy the normality assumption. The Tobit model assumes that the underlying latent outcome follows a Gaussian distribution. Similar to the hurdle model, we can transform the latent variable so that the Gaussian assumption is verified for the observable range of the outcome. For the Tobit model, it is impossible to verify the correct distribution for the latent variable below $L$ where the process is unobserved.


All models were fit using four Markov chains. For the hurdle model, each chain was run for 2,000 iterations, including 750 warm-up iterations. To improve sampler stability and convergence, the Stan control parameters were set to \texttt{adapt\_delta = 0.95} and \texttt{max\_treedepth = 12}. Computation was performed in parallel using four processing cores.

Of note, we fit the separate model (i.e., the hurdle model with \(\rho_0= \rho_1=\rho_2=\rho_3=0\)) was implemented by fitting the continuous and binary components independently using the \texttt{lme4} package: a linear mixed model with Gaussian errors and an identity link for the positive outcomes, and a generalized linear mixed model with a Bernoulli distribution and a probit link for the occurrence process.

\subsection{Relationship between the Tobit and Hurdle Models}

\subsubsection{Notation}

For the formal comparison, we assume a common detection threshold \(L\) and \(Z_{ij} (L)=\mathbb{I}(y_{ij}>L)\). Let
\begin{equation} \label{eq:mean_tobit}
\mu_{ij}^{\text{Tobit}}
=
\beta_0
+
\sum_{k=1}^{K}\beta_k x_{ij}^{(k)}
+
b_{i0}
+
b_{i1}t_{ij}
\end{equation}
denote the conditional mean of the Tobit latent process, given the subject-specific random effects. Similarly, let
\begin{equation} \label{eq:mean_hurdle}
\mu_{ij}^{\text{Hurdle}}
=
\gamma_0
+
\sum_{s=1}^{S}\gamma_s m_{ij}^{(s)}
+
a_{i0}
+
a_{i1}t_{ij}
\end{equation}
denote the conditional mean of the positive-outcome component in the hurdle model.

\subsubsection{Tobit Likelihood}

Under the Tobit model,
\[
y_{ij}^\ast
=
\mu_{ij}^{\text{Tobit}}
+
\varepsilon_{ij},
\qquad
\varepsilon_{ij}\sim N(0,\sigma_\varepsilon^2).
\]
Conditional on \(\boldsymbol b_i\), the probability of exceeding the threshold is
\[
\Pr(y_{ij}>L\mid \boldsymbol b_i)
=
\pi_{ij}(\boldsymbol b_i)
=
\Phi\left(
\frac{\mu_{ij}^{\text{Tobit}}-L}{\sigma_\varepsilon}
\right).
\]
For uncensored observations, define the Gaussian density
\[
f_1(y_{ij}\mid \boldsymbol b_i)
=
\frac{1}{\sigma_\varepsilon\sqrt{2\pi}}
\exp\left\{
-\frac{(y_{ij}-\mu_{ij}^{\text{Tobit}})^2}{2\sigma_\varepsilon^2}
\right\}.
\]
The subject-level likelihood is then
\begin{equation} \label{eq:tobit_conditional_likelihood}
L_i^{\text{Tobit}}
=
\int
\prod_{j=1}^{n_i}
\left[
1-\pi_{ij}(\boldsymbol b_i)
\right]^{1-Z_{ij}(L)}
\left[
f_1(y_{ij}\mid \boldsymbol b_i)
\right]^{Z_{ij}(L)}
f_b(\boldsymbol b_i)
d\boldsymbol b_i,
\end{equation}
where \(f_b(\cdot)\) is the density of \(\boldsymbol b_i\). The full likelihood is
\begin{equation} \label{eq:tobit_full_likelihood}
L^{\text{Tobit}}
=
\prod_{i=1}^{N}
L_i^{\text{Tobit}}.
\end{equation}

\subsubsection{Hurdle Likelihood}

The hurdle model decomposes the observed data distribution into an occurrence component and a positive continuous component. The occurrence component is
\[
\Pr(y_{ij}>L\mid \boldsymbol c_i)
=
\pi_{ij}^{\ast}(\boldsymbol c_i)
=
\Phi(\eta_{ij}),
\]
where
\[
\eta_{ij}
=
\alpha_0
+
\sum_{s=1}^{S}\alpha_s m_{ij}^{(s)}
+
c_{i0}
+
c_{i1}t_{ij}.
\]

For the continuous component, define
\[
f_2(y_{ij}\mid \boldsymbol a_i)
=
\frac{1}{\sigma_e\sqrt{2\pi}}
\exp\left\{
-\frac{(y_{ij}-\mu_{ij}^{\text{Hurdle}})^2}{2\sigma_e^2}
\right\}.
\]
The probability that this Gaussian variable exceeds the threshold is
\[
\pi_{ij}^{\ast\ast}(\boldsymbol a_i)
=
\Pr(y_{ij}>L\mid \boldsymbol a_i)
=
\Phi\left(
\frac{\mu_{ij}^{\text{Hurdle}}-L}{\sigma_e}
\right).
\]
Therefore, the truncated positive density is
\[
g(y_{ij}\mid y_{ij}>L,\boldsymbol a_i)
=
\frac{
f_2(y_{ij}\mid \boldsymbol a_i)
}{
\pi_{ij}^{\ast\ast}(\boldsymbol a_i)
}.
\]
The subject-level hurdle likelihood is
\begin{align} \label{eq:hurdle_likelihood}
\begin{split}
L_i^{\text{Hurdle}}
& =
\iint
\prod_{j=1}^{n_i}
\left[
1-\pi_{ij}^{\ast}(\boldsymbol c_i)
\right]^{1-Z_{ij}(L)}
\left[
\pi_{ij}^{\ast}(\boldsymbol c_i)
\frac{
f_2(y_{ij}\mid \boldsymbol a_i)
}{
\pi_{ij}^{\ast\ast}(\boldsymbol a_i)
}
\right]^{Z_{ij}(L)} \\
& f_{ac}(\boldsymbol a_i,\boldsymbol c_i)
d\boldsymbol a_i d\boldsymbol c_i,
\end{split}
\end{align}
where \(f_{ac}(\cdot,\cdot)\) is the joint density of \(\boldsymbol a_i\) and \(\boldsymbol c_i\). The full likelihood is
\begin{equation} \label{eq:hurdle_full_likelihood}
L^{\text{Hurdle}}
=
\prod_{i=1}^{N}
L_i^{\text{Hurdle}}.
\end{equation}

\subsubsection{Tobit as a Constrained Special Case of the Hurdle Model}

The Tobit model can be recovered from the hurdle model with a probit link under constraints that force the occurrence and positive-outcome components to be governed by the same latent Gaussian process. Specifically, assume common detection limit \(L\), 
\[
K=S, \qquad
x_{ij}^{(k)}=m_{ij}^{(k)} \quad \text{for all } k,
\]
and
\[
\sigma_e=\sigma_\varepsilon,\qquad
\gamma_k=\beta_k \quad \text{for } k=0,\ldots,K,\qquad
\boldsymbol a_i=\boldsymbol b_i.
\] 
With a probit link, impose these constraints
\begin{equation} \label{eq:constraints}
\alpha_0=\frac{\gamma_0-L}{\sigma_e}, \qquad
\alpha_k=\frac{\gamma_k}{\sigma_e} \quad (k=1,\ldots,K),
\qquad
\boldsymbol c_i=\frac{\boldsymbol a_i}{\sigma_e}.
\end{equation}
Then
\[
\eta_{ij}
=
\frac{\mu_{ij}^{\text{Hurdle}}-L}{\sigma_e},
\]
and hence
\[
\pi_{ij}^{\ast}(\boldsymbol c_i)
=
\pi_{ij}^{\ast\ast}(\boldsymbol a_i)
=
\Phi\left(
\frac{\mu_{ij}^{\text{Hurdle}}-L}{\sigma_e}
\right).
\]
For uncensored observations, the density of hurdle model becomes
\[
\pi_{ij}^{\ast}(\boldsymbol c_i)
\frac{
f_2(y_{ij}\mid\boldsymbol a_i)
}{
\pi_{ij}^{\ast\ast}(\boldsymbol a_i)
}
=
f_2(y_{ij}\mid\boldsymbol a_i).
\]
Because the continuous Gaussian process is constrained to equal the Tobit latent Gaussian process, \(f_2(y_{ij}\mid\boldsymbol a_i)=f_1(y_{ij}\mid\boldsymbol b_i)\). The zero contribution also reduces to \(1-\pi_{ij}(\boldsymbol b_i)\). Consequently, the Tobit and hurdle likelihoods in \eqref{eq:tobit_full_likelihood} and \eqref{eq:hurdle_full_likelihood} are identical under these constraints.

The constraint \(\boldsymbol c_i=\boldsymbol a_i/\sigma_e\) collapses the four-dimensional hurdle random-effects structure to a two-dimensional shared random-effects process. Therefore, the exact Tobit special case is a constrained, degenerate form of the general hurdle random-effects model rather than an unconstrained positive-definite four-dimensional Gaussian random-effects model. Without these constraints, the hurdle model allows the probability of exceeding the threshold and the magnitude of positive observations to have distinct fixed effects, variance components, and subject-specific associations.

\subsubsection{Comparison of Model Assumptions}

Both the Tobit and hurdle models can be used for longitudinal semi-continuous outcomes with subject-specific random effects. However, they differ in their interpretation of zero observations and in how the probability of positivity is linked to the positive-outcome distribution.

The Tobit model assumes a single latent Gaussian process. Under this model,
\[
\Pr(y_{ij}>L\mid\boldsymbol b_i)
=
\Phi\left(
\frac{\mu_{ij}^{\text{Tobit}}-L}{\sigma_\varepsilon}
\right).
\]
Thus, covariate effects act on the latent outcome scale and simultaneously influence both threshold crossing and the magnitude of uncensored outcomes. This structure can be efficient when zeros truly arise from censoring at a detection limit, but it may be overly restrictive when zero observations represent true absence or when the data contain more zeros than expected under a latent Gaussian model.

The hurdle model instead specifies separate components for threshold exceedance and positive-outcome magnitude. This allows the same covariate to have different effects on the probability of observing a positive value and on the magnitude of the positive value once the threshold is exceeded. In addition, the joint random-effects covariance structure allows subject-specific association between the occurrence and magnitude processes to be estimated rather than imposed. This flexibility is particularly useful for semi-continuous outcomes with excess zeros, for which the Tobit assumption that all zeros arise from the lower tail of one latent Gaussian process may be inappropriate \citep{moulton1995mixture}. When zeros represent true absence rather than censoring, Tobit estimates can be biased, whereas the hurdle model can provide a more appropriate representation of the data-generating mechanism \citep{kim2009two,noh2019extended}.

\section{Application to Clonal Hematopoiesis} \label{sec:application}

Scientific interest was on studying the environmental and demographic effects on clonal dynamics of mCAs in a large population-based cohort. We analyzed data from the Prostate, Lung, Colorectal, and Ovarian (PLCO) Cancer Screening Trial \citep{andriole2009mortality}.

\citet{kelly2025} identified a subset of 1,490 individuals from a cohort of 56,324 participants. Eligibility was restricted to individuals who i) had a detectable mCA (a positive clonal fraction) at a single cross-sectional examination of the longitudinal process, and ii) had at least two available serial samples. They analyzed clonal fractions (CF) associated with mosaic loss of Y chromosome (mLOY), mosaic loss of X chromosome (mLOX), and autosomal alterations on longitudinal samples that had an associated mosaic alteration. Clonal fractions characterize the proportion of cells in an individual sample that carry a specific mosaic alteration. Changes in clonal fraction indicate that the mosaic alteration process is undergoing clonal expansion or contraction over time. Characterizing the effect of variables such as age and smoking status on the dynamics of clonal fraction may elucidate mechanisms for the role of hematopoiesis on subsequent cancer.  

Linear mixed models for examining exposure and demographic effects on mLOY, mLOX, and autosomal clonal fraction dynamics over a 2-6 year period were fit by \citet{kelly2025}. A major finding was that CF increased with age and that there was a strong interaction between age and smoking for mLOY, such that CF was lower in non-smokers than smokers at younger ages, but the groups were similar at older ages. They only analyzed clonal fractions (positives) and did not consider longitudinal observations without an alteration (zeros). Zero measurements may reflect a true zero, where no abnormality is present, or an undetected mosaic alternation with a very small clonal fraction. The hurdle and Tobit models can be used to accommodate the zero measurements, alleviating concern for potential biased estimation resulting from ignoring these measurements.

We reanalyzed the longitudinal clonal fraction data using the hurdle and Tobit mixed models, both of which incorporate the zero-valued observations while accounting for within-subject correlation. Our goal was to assess the robustness of the findings reported by \citet{kelly2025} and determine whether explicitly modeling the semi-continuous nature of CF leads to different conclusions regarding the effects of age and smoking on mCA progression.



Table~\ref{tab:hurdle_combined} shows the hurdle model results for mLOY and  mLOX alterations. The combined estimates correspond to a hurdle model with a random intercept and slope for both the continuous and binary processes.  The separate estimates correspond to separately fitting a linear mixed model for positive CFs  and a generalized linear mixed model (probit link) for the presence of the particular mosaic alteration (The estimates for the separate linear mixed model  are the same as those presented by \cite{kelly2025}).  These separate estimates are equivalent to fitting the hurdle model under the constraint that there are no associations between random effects from the two processes.

For both  mLOY and mLOX, the hurdle model demonstrates similar estimates between the combined and separate models for the linear mixed model of clonal fraction. Thus, estimates obtained from the hurdle model (with correlated random effects between components) are nearly identical to those present by \cite{kelly2025}, where 
age has a clear positive effect, showing that clonal fraction increases as an individual ages. For mLOY,  there was a strong interaction between age and smoking where biggest differences are seen at younger ages with diminishing differences at older ages. 

In this mosaicism study, longitudinal data on 1,490 individuals of the 56,324 PLCO participants were collected. As previously mentioned, the subsample was chosen based on having at least one mosaic alteration (mLOX, mLOY, or autosomal) on a single cross-sectional evaluation for each participant. For this reason, the binary model parameters are difficult to interpret, but modeling the process  may be important for inference on the continuous outcome of interest, CF. 
That said, we do see increases in the probability of both mLOY and mLOX alterations as age increases. For mLOY, we find that current smokers have the highest CF, followed by former smokers, with never smokers having the lowest CF. 

Table~\ref{tab:hurdle_combined} also shows the correlations between the random effects for intercepts and slopes between the continuous and binary processes, $\rho_0$ and $\rho_1$. Both values are large suggesting that individuals with a large CF are likely to have a higher frequency of mosaic alterations.

Table~\ref{tab:hurdle_autosomal} shows the estimation results from hurdle models for autosomal chromosomes under the combined and separate modeling approaches. Since an individual can have multiple autosomal alterations at a given time point, we expanded our model to include random effects for variation of alterations within individual. We find that autosomal CF increases with age, but there are no effects of smoking. 

The results from Tobit models for mLOX and mLOY (Table~\ref{tab:tobit_combined}) as well as for autosomal alterations (Table~\ref{tab:tobit_autosomal}) were similar to those reported from the corresponding hurdle models. 

\begin{figure}[h]
    \centering
    \includegraphics[width=0.90\linewidth]{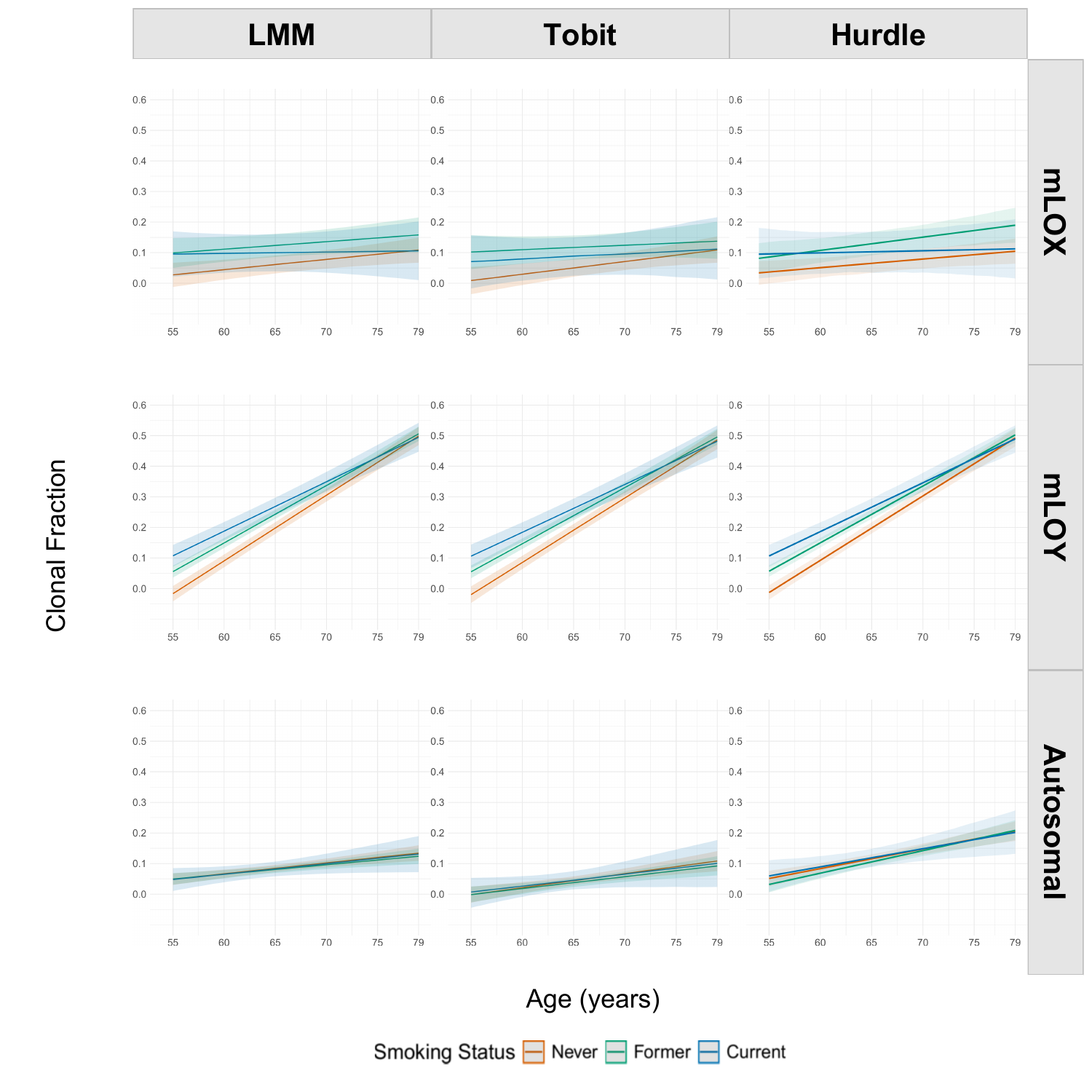}
    \caption{Comparison of coefficient estimates and 95\% confidence intervals from linear mixed models (LMM), hurdle models, and Tobit models for mLOX, mLOY, and autosomal alterations. Rows correspond to outcome variables and columns correspond to modeling approaches.}
    \label{fig:combined_models}
\end{figure}

Although the estimates from \cite{kelly2025} using standard linear mixed models fit only to the positive CF values (i.e., the continuous component of the Separate model) are very similar to those from the Tobit and hurdle models in our application (Figure~\ref{fig:combined_models}), such similarity should not be expected in general.

In the next section, we present simulation results for comparing the different models under a correct and misspecified formulation.

\section{Simulation Results}\label{sec:simulations}





We conducted simulation studies to evaluate the finite-sample performance of the Tobit and hurdle models under correct specification and model misspecification. The primary objective was to assess recovery of fixed-effect, variance, and correlation parameters across three hurdle-generated settings: one satisfying the Tobit-equivalence constraints, one with a larger proportion of zeros than expected under a Tobit model, and one violating the Tobit-equivalence constraints. We further considered Tobit-generated data to compare the Tobit model with a hurdle model and with a linear mixed model fit only to the positive observations, which ignores the censoring mechanism.

Four scenarios were considered. In Scenario I, data were generated from a hurdle model satisfying the constraints in \eqref{eq:constraints}, so that the hurdle model with a probit link was equivalent to a Tobit model. In Scenario II, data were generated from a hurdle model with excess zeros. In Scenario III, data were generated from a hurdle model that violated one Tobit-equivalence constraint by setting the binary time effect to zero, $\alpha_1=0$, while retaining a nonzero continuous time effect, $\gamma_1\neq 0$. In Scenario IV, data were generated from a Tobit model. Scenarios I--III were fit using both hurdle and Tobit models, whereas Scenario IV was fit using Tobit, hurdle, and linear mixed models. The linear mixed model was fit only to the positive continuous observations and therefore ignored censoring.

For simplicity, each simulated dataset consisted of 1{,}000 subjects measured at five equally spaced time points,
\(t_{ij}\in\{0,0.25,0.50,0.75,1\}\).
Each scenario was replicated 1{,}000 times, and the common detection limit was set to \(c=0.7\). Time was the only fixed-effect covariate. The Tobit model included a subject-level random intercept and a residual error term. The hurdle model included subject-level random intercepts in both the binary and positive continuous components, with a residual error term in the continuous component. Hurdle-generated outcomes were semicontinuous, with a point mass at zero and a lognormal distribution for positive values; the probability of a positive outcome was generated using a probit mixed model. Tobit-generated outcomes were generated from a latent lognormal mixed-effects model, exponentiated to the original scale, and left-censored at \(c=0.7\).

Table~\ref{tab:sim_studies_combined} summarizes the results. Panel A shows results for hurdle-generated data in Scenarios I--III, and Panel B shows results for Tobit-generated data in Scenario IV. The table reports the Monte Carlo mean of each estimate, with the mean squared error (MSE) in parentheses.

\subsection{Correctly Specified Models}


The correctly specified analyses are given by the hurdle fit in Scenario I and the Tobit fit in Scenario IV. In Scenario I, where the data were generated from a hurdle model satisfying the Tobit-equivalence constraints, the hurdle model accurately recovered both the continuous-outcome parameters and the binary-process parameters. For example, the estimated means for $\gamma_0$, $\gamma_1$, $\alpha_0$, and $\alpha_1$ were close to their true values, and the variance and correlation parameters were also well recovered.

Similarly, in Scenario IV, when data were generated from a Tobit model and analyzed using the Tobit model, the fixed-effect parameters and variance components were estimated with minimal bias. The mean estimates of $\beta_0$ and $\beta_1$ were nearly identical to their true values, and the estimated random-intercept and residual standard deviations were also close to the data-generating values. These results show that both models perform well when their assumptions match the data-generating mechanism.

\subsection{Tobit-Equivalent and Cross-Model Fits}



Scenario I also illustrates the theoretical relationship between the Tobit and hurdle models. Because the hurdle data-generating mechanism in this scenario was constrained to be equivalent to a Tobit model, the Tobit fit recovered the common parameters well. The Tobit estimates of $\gamma_0$, $\gamma_1$, $\sigma_{a_0}$, and $\sigma_e$ were very close to their true values and comparable to the corresponding hurdle estimates. As expected, the Tobit model did not estimate the separate binary-process parameters $\alpha_0$, $\alpha_1$, $\sigma_{c_0}$, or $\rho_0$.

Conversely, when data were generated from a Tobit model in Scenario IV and analyzed using the hurdle model, the hurdle model also recovered the main Tobit parameters with small bias. This is consistent with the fact that, under a probit link and appropriate parameter constraints, the Tobit model can be viewed as a special case of the hurdle model. Thus, when the equivalence assumptions are satisfied, the two models can produce similar estimates for the common regression and variance parameters.

In contrast, the linear mixed model performed poorly for Tobit-generated data. In Scenario IV, the LMM overestimated the intercept and underestimated the time effect relative to the true Tobit parameters. This bias arises because the LMM analyzes only the observed positive continuous outcomes and ignores the left-censoring mechanism. As a result, the LMM likelihood is misspecified for Tobit data, highlighting the importance of using a model that accounts for censoring when the observed zeros or nondetects arise from a censored latent Gaussian process.

\subsection{Excess-Zero Scenario}

Scenario II evaluates a setting in which zeros arise from a separate occurrence process and are more frequent than would be expected under a Tobit model \citep{moulton1995mixture}. When the hurdle model was fit to these data, it recovered the true parameters accurately. The mean estimates for both the continuous component and the binary component were close to their true values, with relatively small MSEs.

The Tobit model, however, showed substantial bias in this excess-zero setting. In particular, the Tobit fit produced a much more negative estimate of the intercept, an inflated estimate of the time effect, and substantially inflated estimates of the random-intercept and residual standard deviations. This behavior is expected because the Tobit model attributes all zero or nondetect observations to the lower tail of a single latent Gaussian distribution. When zeros are instead generated by a distinct occurrence mechanism, the Tobit likelihood is misspecified, leading to distorted fixed-effect estimates and inflated variance components.

\subsection{Constraint-Violation Hurdle Scenario}

Scenario III examines a hurdle data-generating mechanism that violates one of the Tobit-equivalence constraints. Specifically, the binary time effect was set to $\alpha_1=0$, while the continuous time effect $\gamma_1$ remained nonzero. Under this setting, the hurdle model again recovered the true parameters accurately, including the absence of a time effect in the binary process and the presence of a time effect in the continuous outcome process.

The Tobit model showed noticeable bias in this scenario; the Tobit estimate of the time effect was attenuated relative to the true value, and the variance components were also inflated. This reflects the restrictive nature of the Tobit model: it links the probability of being observed above the detection threshold and the magnitude of the positive outcome through a single latent Gaussian process. When the binary and continuous components follow different regression structures, this assumption is violated.

Overall, the simulation results support the theoretical connection between the Tobit and hurdle models while highlighting important practical differences. When the hurdle model uses a probit link and the Tobit-equivalence constraints hold, the two approaches yield similar estimates for the common parameters. However, when zeros arise from a separate occurrence process, or when the binary and continuous components have different covariate effects, the hurdle model provides more reliable estimation, whereas the Tobit model can yield biased fixed-effect and variance estimates. Conversely, when data are generated from a Tobit model, the hurdle model with a probit link continues to recover the main parameters well. These findings suggest that, when the probit link is appropriate, the hurdle model offers a more flexible framework for semi-continuous outcomes. The results also show that standard linear mixed models fit only to the positive observations, with the censored observations excluded from the analysis, are inappropriate for Tobit-generated data.

\section{Conclusion}\label{sec:discussion}

In this paper, we compared Tobit and two-part hurdle mixed models for semi-continuous longitudinal outcomes, with emphasis on their assumptions, theoretical relationship, and practical performance. Our results show that a Tobit model can be viewed as a constrained special case of a hurdle model with a probit link, provided that the binary occurrence process and the positive continuous process are governed by the same latent Gaussian mechanism. Under these constraints, the probability of exceeding the detection threshold and the distribution of positive outcomes are linked through the same fixed effects, random effects, and residual variance. This relationship clarifies why the two models can yield similar estimates in settings where zeros are consistent with censoring from a common latent process.

The simulation studies supported this theoretical relationship. In the Tobit-equivalent hurdle setting, the Tobit and hurdle models produced similar estimates for the common parameters. Similarly, when data were generated from a Tobit model, the hurdle model with a probit link recovered the main parameters accurately. These findings indicate that the hurdle model with a probit link can reproduce Tobit-like behavior when the Tobit-equivalence constraints are approximately satisfied. The simulations also highlight important practical differences between the two approaches. The Tobit model imposes a strong structural assumption: zeros or nondetects are treated as censored observations arising from the lower tail of the same latent distribution that generates the positive values. This assumption can be efficient and parsimonious when scientifically justified, but it may be inappropriate when zeros represent true absence or arise from a separate occurrence process. In such settings, the Tobit likelihood is misspecified. Consistent with this, the Tobit model showed substantial bias in the excess-zero scenario and noticeable bias when the binary and continuous components had different covariate effects. In contrast, the hurdle model recovered the true parameters well across the hurdle-generated scenarios, reflecting its ability to model the occurrence and positive-outcome processes separately. In practice, it is difficult to empirically verify the assumptions of the Tobit model. Thus, we discourage its use without also fitting a hurdle model as a sensitivity analysis.


In the clonal hematopoiesis application, the Tobit and hurdle models produced broadly similar conclusions across mLOY, mLOX, and autosomal mosaic alterations. Both modeling approaches indicated that clonal fraction increased with age. For mLOY, current smokers had higher clonal fractions, and the age-by-smoking pattern was consistent with previous findings. For mLOX and autosomal alterations, the estimated smoking effects were smaller, whereas age remained positively associated with clonal fraction. The similarity of the Tobit, hurdle, and positive-value linear mixed model estimates in this application suggests that the primary conclusions regarding clonal fraction dynamics were not sensitive to the modeling strategy. However, this agreement should not be interpreted as evidence that positive-only analyses are generally adequate for semi-continuous longitudinal data. In the PLCO application, individuals were selected for longitudinal evaluation on the basis of having a detectable mosaic chromosomal alteration at a cross-sectional assessment. As a result, the binary component of the hurdle model is affected by the study ascertainment process and is less directly interpretable as a model for mCA onset in the underlying PLCO cohort. In this analysis, we therefore treated the zero process primarily as a nuisance component and focused on clonal fraction dynamics among individuals in the positive subpopulation, as in \citep{kelly2025}. The combined hurdle model remains useful in this setting because it incorporates zero-valued longitudinal observations and allows correlation between the random effects governing occurrence and magnitude.

An additional advantage of the combined hurdle model is its ability to estimate the subject-level association between the binary and positive continuous processes. In the clonal hematopoiesis analysis, the estimated cross-component random-effect correlations suggested that individuals with larger clonal fractions tended to have a higher probability of detectable mosaic alterations. Separate modeling approaches, such as fitting a linear mixed model to positive observations and a generalized linear mixed model to occurrence indicators, do not estimate these cross-process correlations directly. When such correlations are present, joint modeling may improve interpretation and protect against bias from imposing independence between the occurrence and magnitude processes.

There are several limitations. First, the Tobit and hurdle models rely on distributional assumptions for the latent or positive continuous outcome. Although transformations can improve the plausibility of Gaussian assumptions for observed positive values, the distribution of the unobserved latent process below the detection threshold in a Tobit model cannot be verified directly from the data. Second, the hurdle model is more flexible than the Tobit model but requires estimation of additional parameters, including separate fixed effects, variance components, and cross-component correlations. 
This can increase computational burden and may require larger sample sizes  in complex longitudinal settings. 
Third, in situations where the data are sparse (e.g.,  small number of zero measurements), estimation for the hurdle model may be unstable. In such a situation, a Tobit model may be the only alternative. 

In summary, the Tobit and hurdle models are closely related but answer subtly different modeling questions. The Tobit model can be viewed as a constrained special case of a hurdle model with a probit link. Furthermore, when zeros are well described as censored observations from a common latent process, the Tobit model provides a parsimonious and interpretable approach. When zeros may arise from a separate occurrence process, or when covariate effects differ between occurrence and magnitude, the hurdle model provides a more flexible and robust framework. 
For semi-continuous longitudinal outcomes such as clonal fraction, the hurdle model is well suited because it accommodates both zero and positive observations and links the occurrence and positive continuous processes through correlated random effects.

\section*{Code and Data Sharing}
The R code for the simulation studies and data analysis can be found on GitHub (\url{https://github.com/Sumaja-Bandreddi/Longitudinal-Semi-Continuous-Data-Analysis}). Access to the PLCO datasets can be requested through the Cancer Data Access System (CDAS) at \url{https://cdas.cancer.gov/}.

\section*{Acknowledgment}
We are grateful to the participants of the Prostate, Lung, Colorectal, and Ovarian (PLCO) Cancer Screening Trial and to the investigators and staff who contributed to the development of this valuable resource. This study uses PLCO datasets of clonal fraction measurements of mosaicism. This work was supported by the Intramural Research Program of the National Institutes of Health (NIH). This work used the computational resources of the NIH High Performance Computing Biowulf cluster (\url{https://hpc.nih.gov}).

\bibliographystyle{unsrtnat} 
\bibliography{refs}

@article{tobin1958estimation,
  title={Estimation of relationships for limited dependent variables},
  author={Tobin, James},
  journal={Econometrica: Journal of The Econometric Society},
  volume={26},
  number={1},
  pages={24--36},
  year={1958},
  publisher={JSTOR}
}

@article{hughes1999mixed,
  title={Mixed effects models with censored data with application to {HIV} {RNA} levels},
  author={Hughes, James P},
  journal={Biometrics},
  volume={55},
  number={2},
  pages={625--629},
  year={1999},
  publisher={Wiley Online Library}
}

@article{olsen2001two,
  title={A two-part random-effects model for semicontinuous longitudinal data},
  author={Olsen, Maren K and Schafer, Joseph L},
  journal={Journal of the American Statistical Association},
  volume={96},
  number={454},
  pages={730--745},
  year={2001},
  publisher={Taylor \& Francis}
}

@article{albert2005modelling,
  title={Modelling longitudinal semicontinuous emesis volume data with serial correlation in an acupuncture clinical trial},
  author={Albert, Paul S and Shen, Joannie},
  journal={Journal of the Royal Statistical Society Series C: Applied Statistics},
  volume={54},
  number={4},
  pages={707--720},
  year={2005},
  publisher={Oxford University Press}
}

@article{noh2019extended,
  title={Extended negative binomial hurdle models},
  author={Noh, Maengseok and Lee, Youngjo},
  journal={Statistical Methods in Medical Research},
  volume={28},
  number={5},
  pages={1540--1551},
  year={2019},
  publisher={SAGE Publications Sage UK: London, England}
}

@article{kim2009two,
  title={Two-part factor mixture modeling: {A}pplication to an aggressive behavior measurement instrument},
  author={Kim, YoungKoung and Muth{\'e}n, Bengt O},
  journal={Structural Equation Modeling},
  volume={16},
  number={4},
  pages={602--624},
  year={2009},
  publisher={Taylor \& Francis}
}

@article{kelly2025,
  title={Longitudinal characterization of mosaic chromosomal alterations identifies factors influencing clonal dynamics of leukocytes},
  author={Kelly, Rebecca L. and Brown, Derek W. and Zhou, Weiyin and Hubbard, Aubrey K. and Young, Corey D. and Barnao, Kara M. and Klein, Alyssa and Dutta, Diptavo and Vogt, Aur{\'e}lie and Liu, Jia and Wang, Jiahui and Huang, Wen-Yi and Freedman, Neal D. and Chanock, Stephen J. and Albert, Paul S. and Machiela, Mitchell J.},
  journal={Nature Communications},
  year={2026},
  note={Under revision}
}

@article{proust2014joint,
  title={Joint latent class models for longitudinal and time-to-event data: a review},
  author={Proust-Lima, C{\'e}cile and S{\'e}ne, Mb{\'e}ry and Taylor, Jeremy MG and Jacqmin-Gadda, H{\'e}l{\`e}ne},
  journal={Statistical methods in medical research},
  volume={23},
  number={1},
  pages={74--90},
  year={2014},
  publisher={SAGE Publications Sage UK: London, England}
}

@article{andriole2009mortality,
  title={Mortality results from a randomized prostate-cancer screening trial},
  author={Andriole, Gerald L and Crawford, E David and Grubb III, Robert L and Buys, Saundra S and Chia, David and Church, Timothy R and Fouad, Mona N and Gelmann, Edward P and Kvale, Paul A and Reding, Douglas J and others},
  journal={New England journal of medicine},
  volume={360},
  number={13},
  pages={1310--1319},
  year={2009},
  publisher={Mass Medical Soc}
}

@article{mullahy1998much,
  title={Much ado about two: reconsidering retransformation and the two-part model in health econometrics},
  author={Mullahy, John},
  journal={Journal of Health Economics},
  volume={17},
  number={3},
  pages={247--281},
  year={1998},
  publisher={Elsevier}
}

@article{farewell2017two,
  title={Two-part and related regression models for longitudinal data},
  author={Farewell, VT and Long, DL and Tom, BDM and Yiu, S and Su, Li},
  journal={Annual Review of Statistics and Its Application},
  volume={4},
  pages={283--315},
  year={2017},
  publisher={Annual Reviews}
}

@article{smith2017marginalized,
  title={A marginalized two-part model for longitudinal semicontinuous data},
  author={Smith, Valerie A and Neelon, Brian and Preisser, John S and Maciejewski, Matthew L},
  journal={Statistical Methods in Medical Research},
  volume={26},
  number={4},
  pages={1949--1968},
  year={2017},
  publisher={SAGE Publications Sage UK: London, England}
}

@article{amemiya1984tobit,
  title={Tobit models: {A} survey},
  author={Amemiya, Takeshi},
  journal={Journal of Econometrics},
  volume={24},
  number={1-2},
  pages={3--61},
  year={1984},
  publisher={Elsevier}
}

@article{liu2019statistical,
  title={Statistical analysis of zero-inflated nonnegative continuous data},
  author={Liu, Lei and Shih, Ya-Chen Tina and Strawderman, Robert L and Zhang, Daowen and Johnson, Bankole A and Chai, Haitao},
  journal={Statistical Science},
  volume={34},
  number={2},
  pages={253--279},
  year={2019},
  publisher={JSTOR}
}

@article{dagne2017joint,
  title={Joint two-part {Tobit} models for longitudinal and time-to-event data},
  author={Dagne, Getachew A},
  journal={Statistics in Medicine},
  volume={36},
  number={26},
  pages={4214--4229},
  year={2017},
  publisher={Wiley Online Library}
}

@article{burkner2017brms,
  title={{brms}: An {R} package for Bayesian multilevel models using {Stan}},
  author={B{\"u}rkner, Paul-Christian},
  journal={Journal of Statistical Software},
  volume={80},
  pages={1--28},
  year={2017}
}

@article{carpenter2017stan,
  title={Stan: A probabilistic programming language},
  author={Carpenter, Bob and Gelman, Andrew and Hoffman, Matthew D and Lee, Daniel and Goodrich, Ben and Betancourt, Michael and Brubaker, Marcus and Guo, Jiqiang and Li, Peter and Riddell, Allen},
  journal={Journal of Statistical Software},
  volume={76},
  pages={1--32},
  year={2017}
}

@article{moulton1995mixture,
  title={A mixture model with detection limits for regression analyses of antibody response to vaccine},
  author={Moulton, Lawrence H and Halsey, Neal A},
  journal={Biometrics},
  volume={51},
   number={4},
  pages={1570--1578},
  year={1995},
  publisher={JSTOR}
}

@article{austin2000use,
  title={The use of the {Tobit} model for analyzing measures of health status},
  author={Austin, Peter C and Escobar, Michael and Kopec, Jacek A},
  journal={Quality of Life Research},
  volume={9},
  number={8},
  pages={901--910},
  year={2000},
  publisher={Springer}
}

@article{chen2016two,
  title={A two-part mixed-effects model for analyzing longitudinal microbiome compositional data},
  author={Chen, Eric Z and Li, Hongzhe},
  journal={Bioinformatics},
  volume={32},
  number={17},
  pages={2611--2617},
  year={2016},
  publisher={Oxford University Press}
}

@article{ren2022semi,
  title={A semi-parametric {B}ayesian model for semi-continuous longitudinal data},
  author={Ren, Junting and Tapert, Susan and Fan, Chun Chieh and Thompson, Wesley K},
  journal={Statistics in Medicine},
  volume={41},
  number={13},
  pages={2354--2374},
  year={2022},
  publisher={Wiley Online Library}
}

@article{finak2015mast,
  title={{MAST}: a flexible statistical framework for assessing transcriptional changes and characterizing heterogeneity in single-cell {RNA} sequencing data},
  author={Finak, Greg and McDavid, Andrew and Yajima, Masanao and Deng, Jingyuan and Gersuk, Vivian and Shalek, Alex K and Slichter, Chloe K and Miller, Hannah W and McElrath, M Juliana and Prlic, Martin and others},
  journal={Genome Biology},
  volume={16},
  number={1},
  pages={278},
  year={2015},
  publisher={Springer}
}

@article{tian2007two,
  title={A two-part model for censored medical cost data},
  author={Tian, Lu and Huang, Jie},
  journal={Statistics in Medicine},
  volume={26},
  number={23},
  pages={4273--4292},
  year={2007},
  publisher={Wiley Online Library}
}

@article{su2009bias,
  title={Bias in 2-part mixed models for longitudinal semicontinuous data},
  author={Su, Li and Tom, Brian DM and Farewell, Vernon T},
  journal={Biostatistics},
  volume={10},
  number={2},
  pages={374--389},
  year={2009},
  publisher={Oxford University Press}
}

@article{cragg1971some,
  title={Some statistical models for limited dependent variables with application to the demand for durable goods},
  author={Cragg, John G},
  journal={Econometrica: Journal of the Econometric Society},
  pages={829--844},
  year={1971},
  publisher={JSTOR}
}

@article{twisk2009longitudinal,
  title={Longitudinal {T}obit regression: a new approach to analyze outcome variables with floor or ceiling effects},
  author={Twisk, Jos and Rijmen, Frank},
  journal={Journal of Clinical Epidemiology},
  volume={62},
  number={9},
  pages={953--958},
  year={2009},
  publisher={Elsevier}
}

\clearpage
\begin{table}
\centering
\caption{Estimation from the hurdle models for mLOX and mLOY alterations under the combined and separate modeling approaches}
\label{tab:hurdle_combined}
\vspace{0.5em}
\footnotesize
\setlength{\tabcolsep}{3pt}
\renewcommand{\arraystretch}{1.1}

\begin{tabular}{lcccc}
\toprule
&
\multicolumn{2}{c}{\textbf{mLOX}}
&
\multicolumn{2}{c}{\textbf{mLOY}}
\\
\cline{2-3}\cline{4-5}
&
\textbf{Combined}
&
\textbf{Separate}
&
\textbf{Combined}
&
\textbf{Separate}
\\
\cline{2-5}
\textbf{Parameter}
&
\multicolumn{4}{c}{\textbf{Estimate (SE)}}
\\
\hline

\multicolumn{5}{l}{\textit{Positive-Outcome Model}} \\

$\gamma_0$ & 0.065 (0.015) & 0.068 (0.015) & 0.197 (0.009) & 0.198 (0.010) \\
$\gamma_1$ & 0.003 (0.001) & 0.003 (0.001) & 0.021 (0.001) & 0.021 (0.001) \\
$\gamma_{2,\mathrm{Former}}$ & 0.065 (0.025) & 0.065 (0.025) & 0.047 (0.012) & 0.045 (0.012) \\
$\gamma_{2,\mathrm{Current}}$ & 0.038 (0.037) & 0.038 (0.035) & 0.070 (0.018) & 0.071 (0.018) \\
$\gamma_{3,\mathrm{Former}}$ & 0.002 (0.002) & 0.002 (0.002) & -0.003 (0.001) & -0.003 (0.000) \\
$\gamma_{3,\mathrm{Current}}$ & -0.002 (0.003) & -0.003 (0.003) & -0.005 (0.002) & -0.005 (0.002) \\

\multicolumn{5}{l}{\textit{Binary Model}} \\

$\alpha_0$ & 1.287 (0.195) & 1.844 (0.369) & 2.683 (0.270) & 4.045 (0.204) \\
$\alpha_1$ & 0.089 (0.030) & 0.167 (0.042) & 0.263 (0.039) & 0.023 (0.034) \\
$\alpha_{2,\mathrm{Former}}$ & 0.619 (0.321) & 0.400 (0.287) & 0.554 (0.256) & 0.176 (0.244) \\
$\alpha_{2,\mathrm{Current}}$ & 0.097 (0.391) & -0.027 (0.395) & 1.059 (0.374) & 0.305 (0.430) \\
$\alpha_{3,\mathrm{Former}}$ & -0.037 (0.043) & -0.070 (0.043) & -0.055 (0.037) & -0.001 (0.041) \\
$\alpha_{3,\mathrm{Current}}$ & 0.001 (0.057) & 0.012 (0.060) & -0.080 (0.055) & 0.016 (0.073) \\

\multicolumn{5}{l}{\textit{Standard Deviations}} \\

$\sigma_{a_0}$ & 1.247 (0.235) & 1.485 (---) & 0.392 (0.005) & 0.391 (---) \\
$\sigma_{a_1}$ & 0.097 (0.041) & 0.099 (---) & 0.100 (0.005) & 0.095 (---) \\
$\sigma_{c_0}$ & 0.162 (0.008) & 0.160 (---) & 1.483 (0.075) & 2.038 (---) \\
$\sigma_{c_1}$ & 0.004 (0.002) & 0.004 (---) & 0.399 (0.035) & 0.335 (---) \\
$\sigma_{e}$ & 0.036 (0.002) & 0.036 (---) & 0.034 (0.001) & 0.040 (---) \\
\multicolumn{5}{l}{\textit{Correlations}} \\

$ \rho_a =\mathrm{cor}(a_{i0},a_{i1})$ & 0.282 (0.309) & 1.000 (---) & 0.248 (0.051) & 0.240 (---) \\
$ \rho_c = \mathrm{cor}(c_{i0},c_{i1})$ & 0.344 (0.232) & 0.410 (---) & 0.019 (0.185) & 0.230 (---) \\
$ \rho_0 = \mathrm{cor}(a_{i0},c_{i0})$ & 0.711 (0.123) & ---  & 0.934 (0.023) & --- \\
$\rho_1 = \mathrm{cor}(a_{i1},c_{i1})$ & 0.492 (0.318) & ---  & 0.348 (0.070) & --- \\
$ \rho_2 = \mathrm{cor}(a_{i0},c_{i1})$ & 0.069 (0.330) & ---  & 0.878 (0.081) & --- \\
$ \rho_3 = \mathrm{cor}(a_{i1},c_{i0})$ & 0.631 (0.256) & ---  & -0.046 (0.154) & ---  \\

\bottomrule
\end{tabular}

\vspace{0.4em}
\begin{minipage}{0.95\linewidth}
\scriptsize
Positive outcomes were modeled as
$y_{ij}\mid y_{ij}>0
=
\gamma_0
+
\gamma_1 t_{ij}
+
\gamma_2 \mathrm{Smoking}_i
+
\gamma_3 \mathrm{Smoking}_i t_{ij}
+
a_{i0}
+
a_{i1}t_{ij}
+
e_{ij},$ 
and the binary component as
$\Pr(y_{ij}>0)
=
\Phi\!\left(
\alpha_0
+
\alpha_1 t_{ij}
+
\alpha_2 \mathrm{Smoking}_i
+
\alpha_3 \mathrm{Smoking}_i t_{ij}
+
c_{i0}
+
c_{i1}t_{ij}
\right).$
Here, $\alpha$ and $\gamma$ denote the fixed-effect coefficients for the binary and positive-outcome components of the hurdle model, respectively. Smoking status is modeled as a three-level categorical variable (never, former, and current smoker), with never smokers serving as the reference group. The coefficients labeled ``Former'' and ``Current'' therefore represent contrasts relative to never smokers. The time variable $t_{ij}$ denotes screening age at the $j$th visit of subject $i$, centered at 65 years, such that $t_{ij}=\mathrm{Age}_{ij}-65$. Standard errors for variance components for the separate model were unavailable because we used the \texttt{lme4} package which does not report them. Here, ``---'' indicates that the parameter was either not estimated by the fitted model or not reported directly by this approach.
\end{minipage}
\end{table}

\clearpage
\begin{table}
\centering
\caption{Estimation from hurdle models for autosomal alterations under the combined and separate modeling approaches}
\label{tab:hurdle_autosomal}
\vspace{0.5em}
\footnotesize
\setlength{\tabcolsep}{3pt}
\renewcommand{\arraystretch}{1.1}

\begin{tabular}{lcc}
\toprule
&
\textbf{Combined}
&
\textbf{Separate}
\\
\cline{2-3}
\textbf{Parameter}
&
\multicolumn{2}{c}{\textbf{Estimate (SE)}}
\\
\hline

\multicolumn{3}{l}{\textit{Positive-Outcome Model}} \\

$\gamma_0$ & 0.115 (0.008) & 0.115 (0.008) \\
$\gamma_1$ & 0.006 (0.001) & 0.006 (0.001) \\
$\gamma_{2,\mathrm{Former}}$ & -0.010 (0.011) & -0.010 (0.011) \\
$\gamma_{2,\mathrm{Current}}$ & 0.004 (0.017) & 0.003 (0.017) \\
$\gamma_{3,\mathrm{Former}}$ & 0.001 (0.001) & 0.001 (0.001) \\
$\gamma_{3,\mathrm{Current}}$ & -0.001 (0.002) & -0.001 (0.002) \\

\multicolumn{3}{l}{\textit{Binary Model}} \\

$\alpha_0$ & 0.344 (0.049) & 0.347 (0.046) \\
$\alpha_1$ & 0.012 (0.008) & 0.006 (0.007) \\
$\alpha_{2,\mathrm{Former}}$ & -0.033 (0.066) & -0.016 (0.062) \\
$\alpha_{2,\mathrm{Current}}$ & 0.074 (0.110) & 0.069 (0.101) \\
$\alpha_{3,\mathrm{Former}}$ & -0.001 (0.010) & -0.008 (0.009) \\
$\alpha_{3,\mathrm{Current}}$ & -0.017 (0.018) & -0.015 (0.017) \\

\multicolumn{3}{l}{\textit{Standard Deviations}} \\

$\sigma_{a_0}$ & 0.433 (0.055) & 0.093 (---)\\
$\sigma_{a_1}$ & 0.044 (0.008) & 0.013 (---) \\
$\sigma_{c_0}$ & 0.088 (0.005) & 0.393 (---) \\
$\sigma_{c_1}$ & 0.014 (0.001) & 0.011 (---)\\
$\sigma_{e}$ & 0.036 (0.002) & 0.031 (---) \\

\multicolumn{3}{l}{\textit{Correlations}} \\

$\rho_a = \mathrm{cor}(a_{i0},a_{i1})$ & -0.295 (0.184) & 1.000 (---) \\
$\rho_c = \mathrm{cor}(c_{i0},c_{i1})$ & 0.108 (0.072) & ---  \\
$\rho_0 = \mathrm{cor}(a_{i0},c_{i0})$ & 0.560 (0.103) & ---  \\
$\rho_1 = \mathrm{cor}(a_{i1},c_{i1})$ & 0.870 (0.075) & ---  \\
$\rho_2 = \mathrm{cor}(a_{i0},c_{i1})$ & -0.032 (0.107) & ---  \\
$\rho_3 = \mathrm{cor}(a_{i1},c_{i0})$ & -0.159 (0.157) & ---  \\

\multicolumn{3}{l}{\textit{Nested Standard Deviations and Correlations}} \\
$\sigma_{a_{ij,0}}$ & 0.334 (0.066) & 0.074 (---) \\
$\sigma_{c_{ij,0}}$ & 0.088 (0.003) & 0.257 (---)\\
$\mathrm{cor}(a_{ij,0},c_{ij,0})$ & 0.345 (0.121) & ---  \\
\bottomrule
\end{tabular}

\vspace{0.4em}
\begin{minipage}{0.95\linewidth}
\scriptsize
Positive outcomes were modeled as
$y_{ijk}\mid (y_{ijk}>0)
=
\gamma_0
+\gamma_1 t_{ij}
+\gamma_2\mathrm{Smoking}_i
+\gamma_3\mathrm{Smoking}_it_{ij}
+a_{i0}
+a_{i1}t_{ij}
+a_{ij,0}
+e_{ijk},$
and the binary component as
$
\Pr(y_{ijk}>0)=\Phi\!\left(
\alpha_0
+\alpha_1 t_{ij}
+\alpha_2 \mathrm{Smoking}_i
+\alpha_3 \mathrm{Smoking}_i t_{ij}
+c_{i0}
+c_{i1}t_{ij}
+ c_{ij,0}
\right).$
Here, $\alpha$ and $\gamma$ denote the fixed-effect coefficients for the binary and positive-outcome components of the hurdle model, respectively. Smoking status is modeled as a three-level categorical variable (never, former, and current smoker), with never smokers serving as the reference group. The coefficients labeled ``Former'' and ``Current'' therefore represent contrasts relative to never smokers. The time variable $t_{ij}$ denotes screening age at the $j$th visit of subject $i$, centered at 65 years, such that $t_{ij}=\mathrm{Age}_{ij}-65$. Random intercepts $c_{ij, 0}$ and $a_{ij, 0}$ account for clustering of observations within mCA events nested within individuals. Certain standard errors are unavailable because the separate-model approach, based on the \texttt{lme4} package, does not report them directly. Here, ``---'' indicates that the parameter was either not estimated by the fitted model or not reported directly by this approach.
\end{minipage}
\end{table}

\clearpage
\begin{table}
\centering
\caption{Estimation from the Tobit models for mLOX and mLOY alterations}
\label{tab:tobit_combined}
\small

\begin{tabular}{lcc}
\toprule
&
\multicolumn{1}{c}{\textbf{mLOX}}
&
\multicolumn{1}{c}{\textbf{mLOY}}
\\
\cline{2-3}
\textbf{Parameter}
&
\multicolumn{2}{c}{\textbf{Estimate (SE)}}
\\
\hline

$\beta_0$ & 0.049 (0.015) & 0.192 (0.010) \\
$\beta_1$ & 0.004 (0.001) & 0.021 (0.001) \\
$\beta_{2,\mathrm{Former}}$ & 0.067 (0.025) & 0.047 (0.012) \\
$\beta_{2,\mathrm{Current}}$ & 0.037 (0.036) & 0.071 (0.018) \\
$\beta_{3,\mathrm{Former}}$ & -0.003 (0.002) & -0.003 (0.001) \\
$\beta_{3,\mathrm{Current}}$ & -0.002 (0.003) & -0.005 (0.002) \\
$\sigma_{b_0}$ & 0.160 (0.008) & 0.158 (0.004) \\
$\sigma_{b_1}$ & 0.002 (0.001) & 0.010 (0.001) \\
$\sigma_{\varepsilon}$ & 0.057 (0.002) & 0.040 (0.001) \\
$\rho_b = \mathrm{cor}(b_{i0}, b_{i1})$ & 0.179 (0.459) & 0.259 (0.058) \\
\bottomrule
\end{tabular}

\vspace{0.5em}
\begin{minipage}{0.95\linewidth}
\footnotesize
The latent outcome was specified as 
\[
y_{ij}^* =
\beta_0 + \beta_1 t_{ij}
+ \beta_2 \mathrm{Smoking}_i
+ \beta_3 \mathrm{Smoking}_it_{ij}
+ b_{i0} + b_{i1} t_{ij} + \varepsilon_{ij},
\]
the observed outcome was generated from the latent outcome as
\[
y_{ij} =
\begin{cases}
0, & y_{ij}^* \le 0,\\
y_{ij}^*, & y_{ij}^* > 0.
\end{cases}
\]
Here, \(\beta\) denotes the fixed-effect coefficients for the Tobit model. Smoking status is modeled as a three-level categorical variable (never, former, and current smoker), with never smokers serving as the reference group. The coefficients labeled ``Former'' and ``Current'' therefore represent contrasts relative to never smokers. The time variable $t_{ij}$ denotes screening age at the $j$th visit of subject $i$, centered at 65 years, such that $t_{ij}=\mathrm{Age}_{ij}-65$. 
\end{minipage} 
\end{table}

\clearpage
\begin{table}
\centering
\caption{Estimation from Tobit model for autosomal alterations}
\label{tab:tobit_autosomal}
\vspace{1em}
\begin{tabular}{lc}
\toprule
\textbf{Parameter} & \textbf{Estimate (SE)} \\
\hline

$\beta_0$ & 0.044 (0.007) \\
$\beta_1$ & 0.005 (0.001) \\
$\beta_{2,\mathrm{Former}}$ & -0.006 (0.010) \\
$\beta_{2,\mathrm{Current}}$ & 0.002 (0.017) \\
$\beta_{3,\mathrm{Former}}$ & -0.001 (0.002) \\
$\beta_{3,\mathrm{Current}}$ & -0.001 (0.003) \\

$\sigma_{b0}$ & 0.075 (0.007) \\
$\sigma_{b1}$ & 0.006 (0.001) \\
$\sigma_{\varepsilon}$ & 0.116 (0.002) \\

$\sigma_{b_{ij,0}}$ & 0.090 (0.004) \\
$ \rho_b = \mathrm{cor}(b_{i0}, b_{i1})$ & 0.488 (0.198) \\

\bottomrule
\end{tabular}

\vspace{0.4em}
\begin{minipage}{0.95\linewidth}
\footnotesize
The latent outcome was specified as 
\[
y_{ijk}^* =
\beta_0 + \beta_1 t_{ij}
+ \beta_2 \mathrm{Smoking}_i
+ \beta_3 \mathrm{Smoking}_it_{ij}
+ b_{i0} + b_{i1} t_{ij} + b_{ij,0} + \varepsilon_{ijk},
\]
the observed outcome was generated from the latent outcome as
\[y_{ijk} =
\begin{cases}
0, & y_{ijk}^* \le 0,\\
y_{ijk}^*, & y_{ijk}^* > 0.
\end{cases}
\]
Here, \(\beta\) denotes the fixed-effect coefficients for the Tobit model. Smoking status is modeled as a three-level categorical variable (never, former, and current smoker), with never smokers serving as the reference group. The coefficients labeled ``Former'' and ``Current'' therefore represent contrasts relative to never smokers. The time variable $t_{ij}$ denotes screening age at the $j$th visit of subject $i$, centered at 65 years, such that $t_{ij}=\mathrm{Age}_{ij}-65$. 
\end{minipage}
\end{table}

\clearpage
\begin{sidewaystable}
\centering
\caption{Simulation results across data-generating and fitting scenarios from 1{,}000 repetitions}
\label{tab:sim_studies_combined}
\scriptsize                       
\setlength{\tabcolsep}{3.5pt}
\renewcommand{\arraystretch}{1.2}

{\scriptsize\textbf{Panel A: Hurdle-generated data}}\\[0.4em]
\begin{tabular}{l *{3}{c c@{\hspace{1pt}}c}}
\toprule
& \multicolumn{3}{c}{Scenario I} & \multicolumn{3}{c}{Scenario II} & \multicolumn{3}{c}{Scenario III} \\
\cmidrule(lr){2-4}\cmidrule(lr){5-7}\cmidrule(lr){8-10}
&      & \multicolumn{2}{c}{Mean (MSE)} &      & \multicolumn{2}{c}{Mean (MSE)} &      & \multicolumn{2}{c}{Mean (MSE)} \\
\cmidrule(lr){3-4}\cmidrule(lr){6-7}\cmidrule(lr){9-10}
Parameter & True & Hurdle & Tobit & True & Hurdle & Tobit & True & Hurdle & Tobit \\
\midrule
$\gamma_{0}$
& 0.500 & 0.499\,(0.030) & 0.501\,(0.030)
& $-0.200$ & $-0.201$\,(0.041) & $-1.597$\,(0.081)
& 0.500 & 0.500\,(0.031) & 0.479\,(0.032) \\
$\gamma_{1}$
& 0.400 & 0.403\,(0.026) & 0.398\,(0.025)
& 1.000 & 1.005\,(0.058) & 1.348\,(0.081)
& 0.400 & 0.404\,(0.027) & 0.288\,(0.027) \\
$\alpha_{0}$
& 1.428 & 1.447\,(0.076) &  ---
& 0.700 & 0.705\,(0.035) & ---
& 1.428 & 1.439\,(0.072) & --- \\
$\alpha_{1}$
& 0.667 & 0.657\,(0.083) &  ---
& 0.500 & 0.500\,(0.036) & ---
& 0.000 & $<$0.001\,(0.074) & --- \\
$\sigma_{a_0}$
& 0.800 & 0.800\,(0.022) & 0.804\,(0.021)
& 0.400 & 0.401\,(0.035) & 1.304\,(0.054)
& 0.800 & 0.800\,(0.023) & 0.873\,(0.023) \\
$\sigma_{c_0}$
& 1.333 & 1.359\,(0.067) & ---
& 0.800 & 0.797\,(0.024) & ---
& 1.333 & 1.355\,(0.062) & ---\\
$\rho_{0}$
& 1.000 & 0.985\,(0.010) & ---
& 0.000 & 0.003\,(0.056) & ---
& 1.000 & 0.986\,(0.009) & --- \\
$\sigma_{e}$
& 0.600 & 0.599\,(0.007) & 0.599\,(0.007)
& 0.600 & 0.599\,(0.010) & 1.455\,(0.033)
& 0.600 & 0.599\,(0.008) & 0.640\,(0.008) \\
\bottomrule
\end{tabular}

\vspace{1.2em}

{\scriptsize\textbf{Panel B: Tobit-generated data}}\\[0.4em]
\begin{tabular}{l c c c c}
\toprule
& & \multicolumn{3}{c}{Scenario IV} \\
\cmidrule(lr){3-5}
& & \multicolumn{3}{c}{Mean (MSE)} \\
\cmidrule(lr){3-5}
Parameter & True & Tobit & LMM & Hurdle \\
\midrule
$\beta_{0}$            & 0.500 & 0.501\,(0.030) & 0.746\,(0.024) & 0.504\,(0.033) \\
$\beta_{1}$            & 0.400 & 0.398\,(0.025) & 0.313\,(0.024) & 0.396\,(0.029) \\
$\sigma_{b_0}$         & 0.800 & 0.804\,(0.021) & 0.597\,(---)     & 0.802\,(0.024) \\
$\sigma_{\varepsilon}$ & 0.600 & 0.599\,(0.007) & 0.547\,(---)     & 0.598\,(0.009) \\
\bottomrule
\end{tabular}

\vspace{0.4em}
{\par\footnotesize\raggedright
Mean is the average estimate over 1{,}000 replications; MSE, shown in parentheses, is the mean squared error. 
``---'' indicates either that the parameter was not estimated by the fitted model or, for variance-component estimates from the LMM fit using \texttt{lme4}, that the corresponding MSE was not reported. 
$\rho_{0}=\mathrm{cor}(a_{i0},c_{i0})$. 
Scenarios I--III correspond to hurdle-generated data: (i) a hurdle model satisfying the constraints in \eqref{eq:constraints}, 
(ii) an excess-zero hurdle model, and (iii) a hurdle model violating one constraint by setting $\alpha_{1}=0$ while $\gamma_1\neq0$. 
Each was fit using both hurdle and Tobit models. Scenario IV corresponds to Tobit-generated data fit using Tobit, LMM, and hurdle models. \par}
\end{sidewaystable}

\clearpage






\end{document}